\documentclass[a4paper, amsfonts, amssymb, amsmath, reprint, nofootinbib, twoside]{revtex4-1}
\usepackage[left=1.8cm, right=1.8cm, top=1.8cm, bottom=2.0cm]{geometry}
\usepackage[english]{babel}
\usepackage[utf8]{inputenc}
\usepackage[colorinlistoftodos, color=green!40, prependcaption]{todonotes}
\usepackage{amsthm}
\usepackage{amsmath} 
\usepackage{mathtools}
\usepackage{physics}
\usepackage{xcolor}
\usepackage{graphicx}
\usepackage{adjustbox}
\usepackage{placeins}
\usepackage[T1]{fontenc}
\usepackage{lipsum}
\usepackage{csquotes}
\usepackage[colorlinks,allcolors=blue!50!black]{hyperref} 
\makeatletter
\renewcommand\frontmatter@abstractwidth{12cm}
\makeatother

\newcommand{\plus}{\scalebox{0.6}{$+$}}
\newcommand{\ssim}{\raisebox{0.8pt}{\scalebox{0.7}{$\sim$}}}
\newcommand{\angstrom}{\textup{\AA}}

\begin{document}

\title{Materials Informatics Reveals Unexplored Structure Space in Cuprate Superconductors}

\author{Rhys E. A. Goodall}
    \affiliation{Cavendish Laboratory, University of Cambridge, Cambridge CB3 0HE, UK.}
    
\author{Bonan Zhu}
    \affiliation{Department of Materials Science \& Metallurgy, University of Cambridge, Cambridge CB3 0FS, UK}
    
\author{Judith L. MacManus-Driscoll}
    \affiliation{Department of Materials Science \& Metallurgy, University of Cambridge, Cambridge CB3 0FS, UK}

\author{Alpha A. Lee}
     \email[Correspondence email address: ]{aal44@cam.ac.uk}
    \affiliation{Cavendish Laboratory, University of Cambridge, Cambridge CB3 0HE, UK.}

 
\begin{abstract}
High-temperature superconducting cuprates have the potential to be transformative in a wide range of energy applications. 
In this work we analyse the corpus of historical data about cuprates using materials informatics and re-examine how their structures are related to their critical temperatures (Tc).
The available data is highly clustered and no single database contains all the features of interest to properly examine trends. 
To work around these issues we employ a linear calibration approach that allows us to utilise multiple data sources -- combining fine resolution data for which the Tc is unknown with coarse resolution data where it is known. 
The hybrid data set constructed enables us to explore the trends in Tc with the apical and in-plane copper-oxygen distances. 
We show that large regions of the materials space have yet to be explored and highlight how novel experiments relying on nano-engineering of the crystal structure may enable us to explore such new regions. 
Based on the trends identified we propose that single layer Bi-based cuprates are good candidate systems for such experiments.
\end{abstract}


\maketitle

\section*{Introduction}

Since the discovery of high-temperature superconductivity (HTS) in cuprate materials in 1986 \cite{bednorz1986possible}, significant amounts of research has been devoted to understanding, tuning, doping, and growing new cuprates to understand and optimise their properties -- both their critical transition temperatures (Tc) and other important properties such as the coherence length of the superconducting charge carriers. 
\textcolor{black}{This work has uncovered a variety of scaling laws \cite{homes2004universal, uemura1989universal} and structure-function relationships \cite{ohta1991apex, rao_structureproperty_1995, macmanus-driscoll_future_2011} that provide insight into the origin of superconductivity in cuprates and other superconducting systems. However, 
the many different variables cannot all be optimised at the same time, and furthermore, some of them are interdependent, making it difficult to establish causation vs correlation for these trends. In particular, bond distances cannot be independently tuned in the 3 orthogonal directions. The most common way to tune the structure is via thin film epitaxy where the substrate tunes the in-plane lattice parameter. However, the out-of-plane lattice parameter is not fixed and it responds elastically to the in-plane strain. Recently, experimental techniques based on nano-engineering \cite{choi_3d_2019, choi2020interfaces} have been reported that enable independent tuning of lattice distances. These techniques open up new experimental pathways to increase Tc through the exploitation of structure-function relationships.}
Several structure-function relationships are now well established in the literature \textcolor{black}{for HTS in cuprates}. The most important factors believed to relevant for the Tc are: 
i) The concentration of charge carriers in the conduction planes, 
ii) The nature of bonding in the charge-reservoir layers, 
iii) The in-plane Cu-O distance, and
iv) The apical Cu-O distance (here we do not consider electron-doped superconducting cuprates without apical oxygens i.e. those adopting a T' structure). 


In this work we focus on the structure-dependent relationships between Tc and the apical and in-plane Cu-O distances.
The influence of the apical Cu-O distance on Tc is attributed to its effect on the localisation of charge carriers in the CuO\textsubscript{2} planes \textcolor{black}{\cite{pavarini2001band}}. 
The importance of the in-plane Cu-O distance is believed to be due to its impact on Cu-O-Cu super-exchange in the CuO\textsubscript{2} planes.
Such attempts to couple experimentally observed trends to physically relevant mechanisms provides qualitative understanding and insight into the nature of superconductivity in these systems.
Unfortunately, \textcolor{black}{whilst progress has been made on theories describing the percolative nature through which superconductivity emerges in cuprates \cite{popvcevic2018percolative, pelc2018emergence, pelc2019universal} and the importance of inhomogeneity in such processes}, the consolidation of satisfactory quantitative theories for HTS in cuprates \textcolor{black}{that reflect known structure-function relationships and are capable of making predictions about Tc} remains a challenge \cite{plakida2010high, fradkin2015colloquium}.

In the absence of an accepted mechanistic theory, the potential availability of large amounts of historical data and high impact applications has led some researchers towards data-driven phenomenological approaches i.e. machine learning.
Thus far, most work in this area has focused on building models for predicting Tc \textcolor{black}{given a set of easy to evaluate descriptors that represent the materials in question} \cite{stanev2018machine, konno2018deep, liu2020material}.
The hope is that such models may enable the discovery of new families of high-temperature superconductors by first detecting abstract empirical patterns in featurisations of materials currently known to display superconductivity, and then screening new materials based on similarity to these identified patterns.
However, questions exist about whether such approaches will be fruitful when tested experimentally as the evaluation metrics used in proof of concept workflows are often not reflective of real materials discovery workflows \cite{meredig2018can}.
A less explored but similar avenue is how materials informatics approaches can be used in conjunction with careful physical insights to probe our understanding of systems already known to display superconductivity \cite{kim2018apical}.
It is this avenue of enquiry we pursue in this work.

\textcolor{black}{By combining high-resolution data structural data where the Tc is unknown with coarse-resolution data where Tc is known, we show the existence of unexplored regions of the materials space defined by the apical and in-plane Cu-O distances that are ripe for further experimental investigation. Our approach focuses solely on the apical and in-plane Cu-O distances as the limited availability of data precludes direct inclusion of other important factors know to affect the physics. However, by selecting the points with highest Tc given the cuprate family in different regions of the lattice parameter space we are able to restrict ourselves to points where the other physical parameters are likely to be near their optima. Our results highlight how materials informatics can play an important role in helping to guide experimental efforts in material science.}

\section*{Inferring structural parameters of cuprates}

\begin{figure*}[!t]
\centering
\includegraphics[width=17cm]{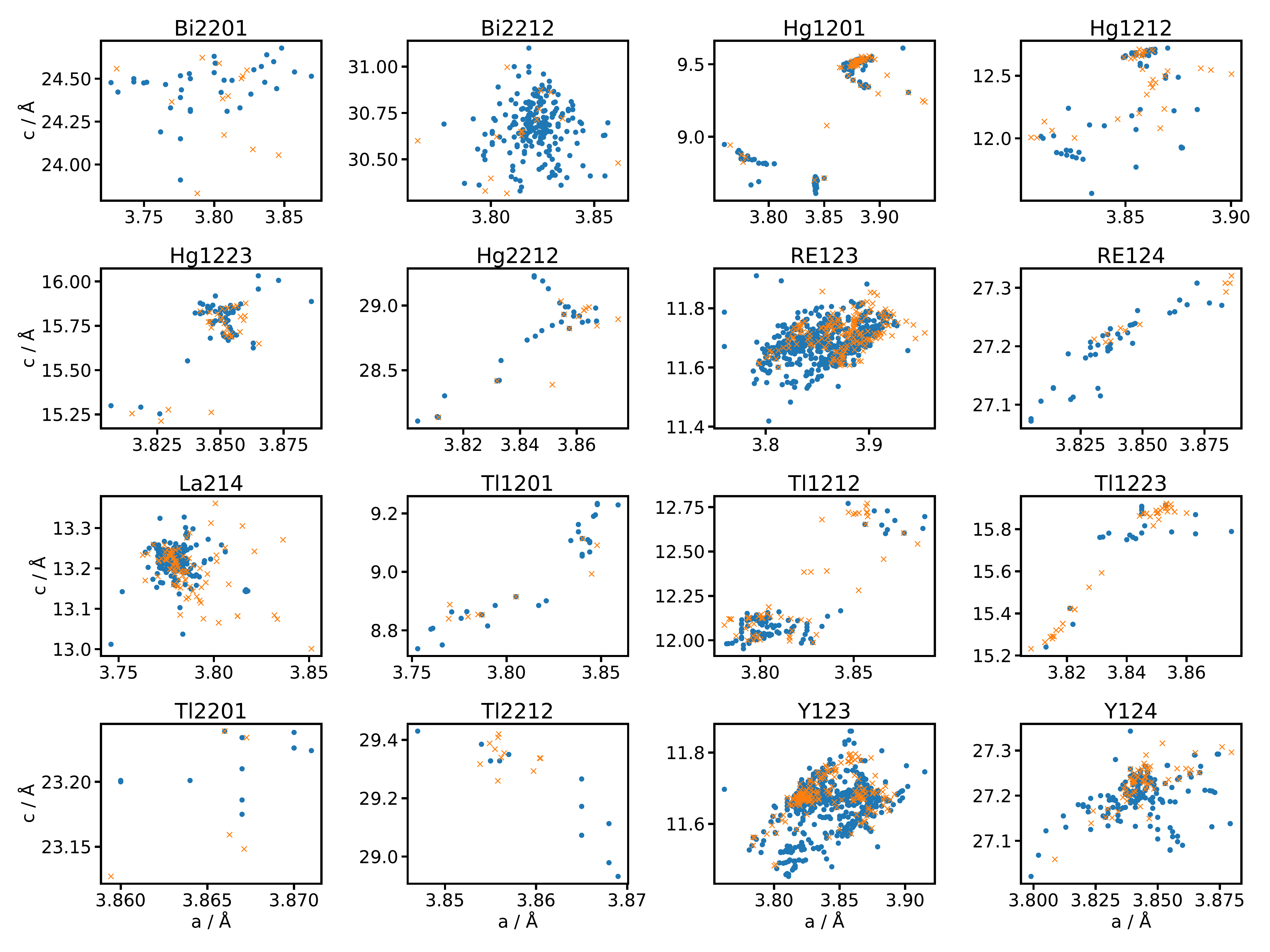}
\caption{The figure shows the overlap in the lattice parameters between the SuperCon data set (blue dots) and the ICSD reference data (orange crosses). See \autoref{tab:abbrev} for a list of abbreviations.}
\label{fig:overlap}
\end{figure*}

The principal data source for this work is the \mbox{\textit{SuperCon}} database compiled and distributed by the Japanese National Institute for Material Science. 
Whilst \textit{SuperCon} records the critical temperatures of an extensive range of superconducting materials the information available for each composition is minimal -- structural information is only available for a small proportion of the entries and is limited to the lattice parameters when available.
Unfortunately, it is the common structure shared between different cuprates, characterised by the apical and in-plane distances of the superconducting CuO\textsubscript{2} planes, that is interesting for examining trends.

Whilst the lattice parameters can be determined relatively easily via x-ray diffraction, directly measuring the atomic positions, needed to determine the apical and in-plane distances, typically involves much more specialised x-ray diffraction apparatus or neutron diffraction experiments.
Fortunately, many cuprate structures are already recorded in the Inorganic Crystal Structure Database (ICSD) \cite{hellenbrandt2004inorganic}.
However, the critical temperatures of these materials are not recorded alongside their structures, therefore, limiting the utility of ICSD as a data source for studying Tc-structure trends. 

Consequentially, whilst large amounts of data are available on cuprates, the structure of that data is incomplete in terms of the information required to look at structure-function relationships.
As a result, trends are often examined between relatively small numbers of selected data-points that contain the necessary information for the analysis of a given structure-function hypothesis.
Selection of data in this way has the potential to lead to overconfidence in Tc-structure trends observed.
Below we attempt to overcome this issue of incomplete data by obtaining estimates for the apical and in-plane distances from the knowledge of the more readily available lattice parameters.
Whilst the accuracy of such an approach is diminished and prevents truly quantitative analysis, using estimates allows for a far greater number of examples to be considered, therefore ensuring the robustness of any trends that remain.

For cuprates, the a-lattice parameter is closely related to the in-plane distance.
Assuming that the CuO\textsubscript{2} planes are approximately square planar this entails that the in-plane distance can be estimated as a$/2$ for tetragonal phases and a$/2\sqrt{2}$ for octahedral phases.
This assumption breaks down for cuprates under pressure where the CuO\textsubscript{2} planes generally tend to buckle to relieve pressure on the structure. 

\begin{table}[]
\caption{\textcolor{black}{Abbreviations used to describe the cuprate systems explored in this work. We make use of the A-jk(n-1)n four-digit notation for cuprates of the form A\textsubscript{j}B\textsubscript{k}S\textsubscript{n-1}Cu\textsubscript{n}O\textsubscript{j+k+2n} described in \cite{poole2014super} for Tl, Hg and Bi based cuprates. In the RE123 and RE124 abbreviations RE denotes a range of rare earth metals i.e. RE=\{Nd, Gd, Pr, etc.\}. The representative formulas given are not exclusive and the materials contained in both SuperCon and ICSD contain a variety of dopants e.g. Y for Ca in Bi2212 to reduce cation disorder \cite{eisaki2004effect} and elemental substitutions e.g. Sr for Ba in Hg1201 to produce Ba-free cuprates \cite{hyatt1998synthesis}.}}
\label{tab:abbrev}
\begin{tabular}{lc}
\hline
Family Label & Representative Formula \\ \hline
La214 (T) & La\textsubscript{2}CuO\textsubscript{4}/La\textsubscript{2-x}Sr\textsubscript{x}CuO\textsubscript{4} \\
Y123 & YBa\textsubscript{2}Cu\textsubscript{3}O\textsubscript{7} \\
RE123 & YBa\textsubscript{2}Cu\textsubscript{3}O\textsubscript{7} \\
Y124 & YBa\textsubscript{2}Cu\textsubscript{4}O\textsubscript{8} \\
RE124 & YBa\textsubscript{2}Cu\textsubscript{4}O\textsubscript{8}, \\
Tl1201 & TlBa\textsubscript{2}CuO\textsubscript{5}\\
Tl1212 & TlBa\textsubscript{2}CaCu\textsubscript{2}O\textsubscript{7} \\
Tl1223 & TlBa\textsubscript{2}Ca\textsubscript{2}Cu\textsubscript{3}O\textsubscript{9} \\
Tl2201 & Tl\textsubscript{2}Ba\textsubscript{2}CuO\textsubscript{6} \\
Tl2212 & Tl\textsubscript{2}Ba\textsubscript{2}CaCu\textsubscript{2}O\textsubscript{8}\\
Hg1201 & HgBa\textsubscript{2}CuO\textsubscript{5}\\
Hg1212 & HgBa\textsubscript{2}CaCu\textsubscript{2}O\textsubscript{6} \\
Hg1223 & HgBa\textsubscript{2}Ca\textsubscript{2}Cu\textsubscript{3}O\textsubscript{9} \\
Hg2212 & Hg\textsubscript{2}Ba\textsubscript{2}YCu\textsubscript{2}O\textsubscript{8} \\
Bi2201 & Bi\textsubscript{2}Sr\textsubscript{2}CuO\textsubscript{6} \\
Bi2212 & Bi\textsubscript{2}Sr\textsubscript{2}CaCu\textsubscript{2}O\textsubscript{8}\\ \hline
\end{tabular}
\end{table}

\begin{figure*}[t]
\centering
\includegraphics[width=17cm]{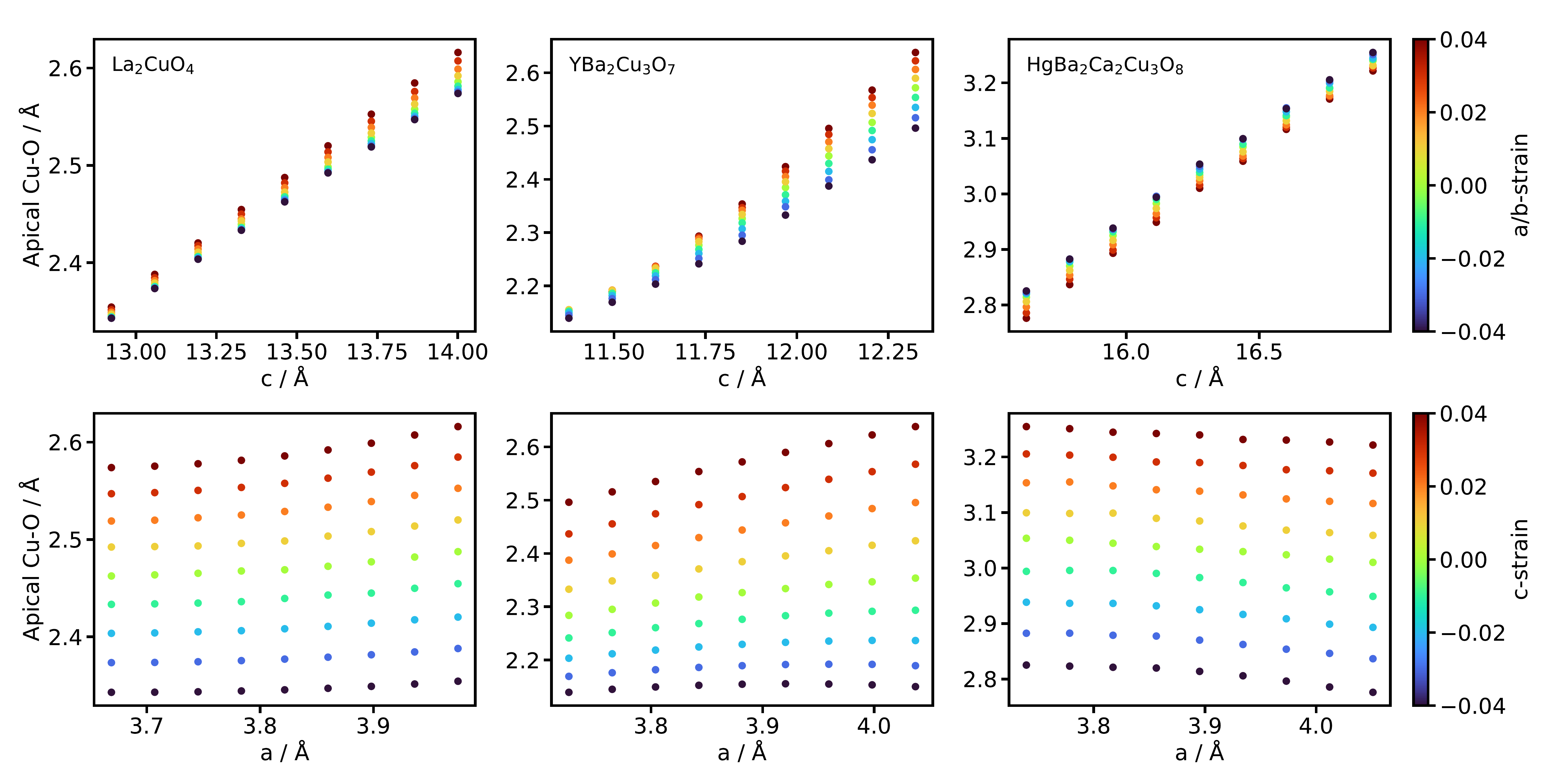}
\caption{The figure shows how the apical distance, as calculated via density functional theory, changes for three prototype cuprate systems that have strained along the c axis and in the a/b plane.
The plots show broadly linear trends for the behaviour of the apical distance as strains are applied.
This result validates the use of linear models to estimate the apical distance from experimental lattice measurements.}
\label{fig:dft}
\end{figure*}

Whilst the c-lattice parameter is often considered as a proxy for the apical distance there is little correlation between the two -- the c-lattice parameter also depends on the thickness of the charge-reservoir layers which can vary significantly between different families of cuprates. 
Consequentially, to estimate the apical distances each family has to be treated independently. The approach adopted here is to use linear calibration models constructed on reference data that relate the a and c-lattice parameters to the apical distance.
Figure \ref{fig:overlap} shows a scatter plot of the structure space as characterised by the a and c-lattice parameters for both the source data (\textit{SuperCon}) and the reference data (ICSD).
The 16 cuprate families investigated here were selected due to having greater than 5 examples in both ICSD and \textit{SuperCon}.
We see that both data sets assign density to similar regions of the structure space for these families.

The variation in the apical distance within families is due to the chemical pressure that arises from doping or atomic substitutions.
The simplest model is that the pressure imparts a uniform stress along the c axis that strains the material.
If the total strains are in the Hookean regime the strain along the c-axis should then be directly proportional to the strain in the apical distance - the implication being that we can approximate the materials' layered structure with hypothetical slabs of constant Young's modulus.
The strains along a and c will also be related by the Poisson effect.
Therefore, the minimal linear calibration model for the apical distances, $\hat{d}_{\text{apical}}$, that also includes this effect is:
\begin{equation}\label{eq:calib}
    \hat{d}_{\text{apical}} = \alpha c + \beta a^* + \gamma
\end{equation}
Where $\alpha, \beta$ and $\gamma$ are the parameters of the calibration model that need to be fitted for each family.
In each case a robust linear model based on the Huber penalty \cite{huber1992robust} was used to reduce the effect of outliers when fitting the calibration models of the form \eqref{eq:calib}.
Models were fitted for the following families: La214 (T) CuO\textsubscript{4}, Y123, RE123, Y124, RE124, Tl1201, Tl1212, Tl1223, Tl2201, Tl2212, Hg1201, Hg1212, Hg1223, Hg2212, Bi2201, and Bi2212 (See \autoref{tab:abbrev} for a list of abbreviations).
To increase the amount of reference available data both neutron and x-ray scattering structures recorded in ICSD were used - ideally only structures derived from neutron scattering would be used as the oxygen positions for x-ray derived structures can be affected by systematic errors.
However, here we believe the increased abundance and diversity of reference data outweights this potential loss of accuracy.

\begin{figure*}[t]
\centering
\includegraphics[width=17.8cm]{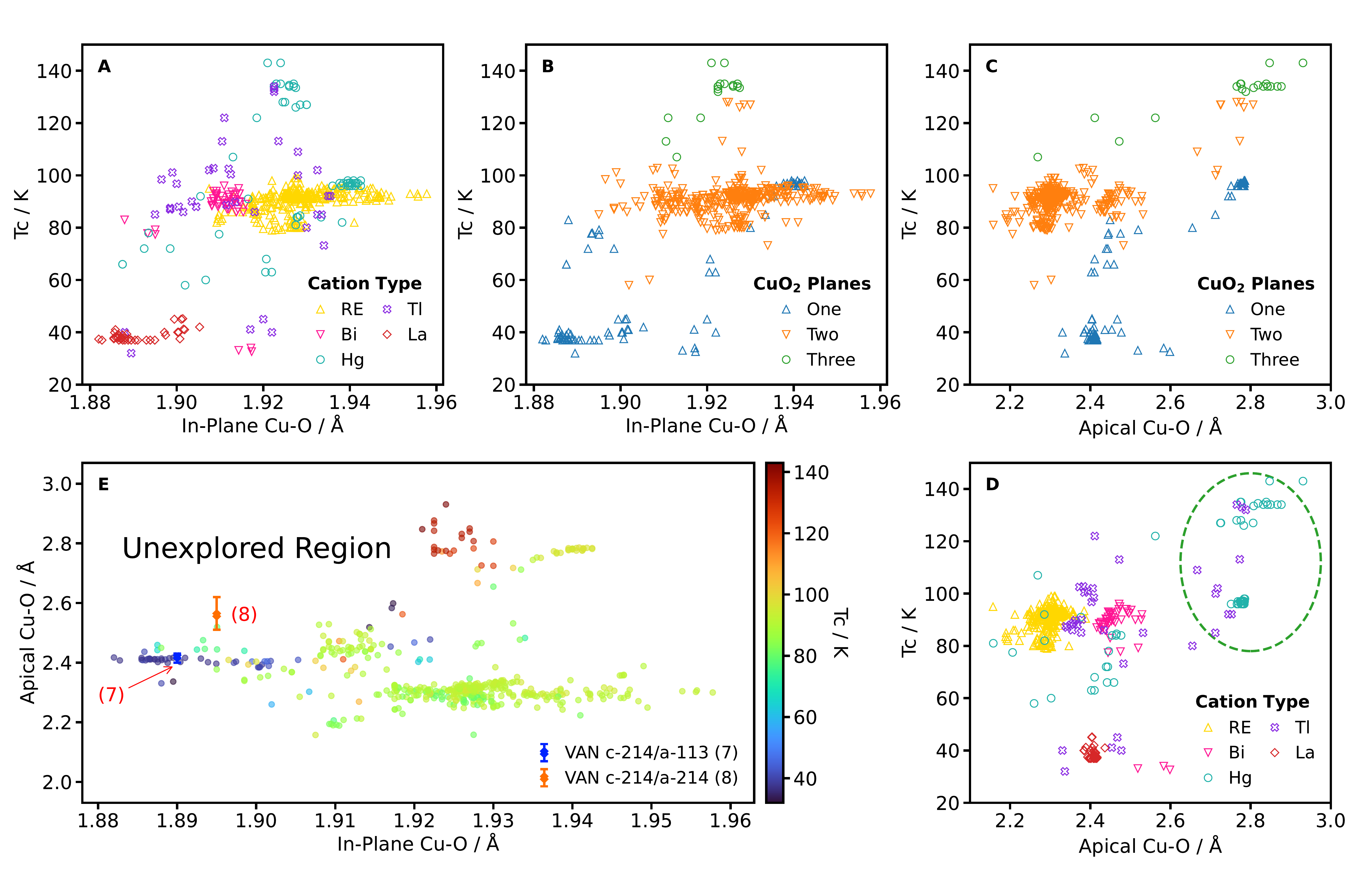}
\caption{Panels \textbf{A-D} show the trends between the critical temperature and the apical and in-plane distances stratified by the number of CuO\textsubscript{2} planes and cation type.
In panel \textbf{A} and \textbf{B} we see the apparent optimium in the in-plane distance of ~1.92 {\AA} for Hg and Tl-based materials.
Panels \textbf{B} and \textbf{C} clearly highlight the trend that Tc increases with the number of CuO\textsubscript{2} planes.
The green circled region in \textbf{D} shows that the trend of Tc increasing with apical distance is only apparent due to Hg and Tl-based materials with high apical distances.
\textbf{E} shows the variation of Tc with the apical and in-plane distances.
A large region (labelled ``Unexplored Region'') of high apical Cu-O distance and low in-plane Cu-O distance is apparent.
Experiments that probe this region are likely to provide useful insight into the nature of Tc-structure trends in cuprates. \textcolor{black}{We highlight the vertically aligned nanocomposite (VAN) samples from \cite{choi_3d_2019, choi2020interfaces} as examples of nano-engineering approaches to investigate the region. In the key VAN c-214/a-113 refers to the La\textsubscript{2}CuO\textsubscript{4-$\delta$}/LaCuO\textsubscript{3} interface from \cite{choi_3d_2019} and VAN c-214/a-214 refers to the c-aligned La\textsubscript{2}CuO\textsubscript{4-$\delta$}/a-aligned La\textsubscript{2}CuO\textsubscript{4-$\delta$} interface from \cite{choi2020interfaces}.}}
\label{fig:groups}
\end{figure*}

To check the validity of this simple approach we employ density functional theory (DFT) to investigate how the apical Cu-O distance changes as the lattice is strained.
We only use DFT as a proxy to look for qualitative trends that are likely to be mirrored in real systems.
This is due to well-documented discrepancies in the lattice parameters between the structures of cuprates as reconstructed from neuron/x-ray scattering experiments and the relaxed structures returned from DFT calculations.

We take La\textsubscript{2}CuO\textsubscript{4}, YBa\textsubscript{2}Cu\textsubscript{3}O\textsubscript{7}, HgBa\textsubscript{2}Ca\textsubscript{2}Cu\textsubscript{3}O\textsubscript{8} as prototypical test cases being illustrative of one, two and three-layer cuprates respectively.
Figure \ref{fig:dft} shows that for tensile and compressive strains of up to 4\% the responses show that monotonic and broadly linear trends exist between the a and c-lattice parameters and the apical distance for the prototype systems explored.
The dependence on the c-lattice parameter is stronger than on the a-lattice parameter as expected.
As the strains become larger some degree of non-linearity does appear, however, the maximal strains examined here are significantly larger than the spread in the experimental data.
Whilst a more complicated model could be used to fit this non-linearity in the clean data obtained via DFT for the experimental data other factors such as systematic variations between different experimental setups cannot be accounted for.
Therefore, adding additional terms to the calibration models without strong physically-motivated priors for their inclusion is undesirable.

\section*{Materials informatics reveals unexplored structure space}

Beyond the apical and in-plane distances there are other important factors known to influence Tc that need to be considered.
Unfortunately, many of these are typically harder to quantify, for example, the nature of bonding in the charge-reservoir layers.
Perhaps the most important of these factors is that achieving optimal oxygen-doping is necessary to maximise Tc.
Given the aim of maximising Tc we are generally only interested in trends between materials characterised in optimal states.
Unfortunately, materials in \textit{SuperCon} are commonly reported with unknown oxygen concentrations.
This is problematic because achieving optimal oxygen doping for a given composition within a family depends on which/whether other dopants are present.
As a result naively selecting the materials with the highest Tcs would end up discarding much of the diversity in the data set making it difficult to establish trends.
To ensure that we maintain as much structural diversity as possible we first perform k-means clustering \cite{lloyd1982least} on the data in the a and c-lattice parameter space and then take the top 20\% of each cluster by Tc.
\textcolor{black}{This selection strategy ensures that the data points considered in the subsequent analysis are diverse in terms of their apical and in-plane Cu-O distances but are also likely to be close to optimal in terms of the other relevant factors for optimising Tc}.
Stratifying the remaining data set into different sub-groups we observe the following trends:

\begin{enumerate}
    \itemsep0em 
    
    \item If cuprate materials are differentiated according to the number of CuO\textsubscript{2} planes in the unit cell there is a clear separation between the different groups (Figures \ref{fig:groups}B and \ref{fig:groups}C).
    The observed increases in Tc with the number of planes are well understood considering intralayer interactions between CuO\textsubscript{2} planes \cite{di2018superconductivity}. \textcolor{black}{A similar separation is also clearly visible in the Uemura relation \cite{uemura1989universal} where the saturation and suppression of the Tc occurs at different relaxation rates depending on the number of CuO\textsubscript{2} planes.}
    
    \item Whilst grouping by the number of CuO\textsubscript{2} planes emphasises a strong positive correlation observed between Tc and the apical distance (Figure \ref{fig:groups}C), grouping materials via the main cation (Figure \ref{fig:groups}D) shows that all the highest Tcs come from Hg and Tl-based materials with high apical distances (Circled in green in Figure \ref{fig:groups}D).
    We note that higher critical temperatures have been achieved in both two and three layer Hg-based cuprates via the application of hydro-static pressure \cite{gao1994superconductivity, yamamoto2015high} which is known to decrease the apical distance \cite{armstrong1995crystal} with minimal impact on the in-plane distance. However, these increases in Tc have been attributed to the effect of pressure on the position of Ba atoms in the buffer layer \cite{gatt1998pressure, volkova2000correlation}. 
    
    \item There is an apparent optimum in-plane distance for Hg and Tl-based cuprates around $\ssim 1.92\angstrom$ (Figures \ref{fig:groups}A).
    In contrast, Rare-Earth-based (RE) cuprates show very little variation in Tc as the in-plane distance changes.
    Looking at the highest Tc Bi-based materials, there is a slight increase in Tc as the in-plane distance approaches  $1.92\angstrom$ but as there are no Bi-based materials reported with in-plane distances above $1.92\angstrom$ in the data sets examined, it is not apparent whether a drop off in Tc, as is the case for Hg and Tl-based materials, would be observed.
    The slight increase in Tc with the in-plane distance for these Bi-based materials could perhaps be attributed to changes in the multi-layer structure between the high Tc Bi2201 type materials (Bi\textsubscript{2\plus x}Sr\textsubscript{2-x-y}Ca\textsubscript{y}CuO\textsubscript{6\plus $\delta$}) \cite{yoshizaki2007substitution} and the Bi2212 (Bi\textsubscript{2}Sr\textsubscript{2}CaCu\textsubscript{2}O\textsubscript{8\plus $\delta$}) family.
    This suggests that for constant apical distance there may be no strong dependence on the in-plane distance for the Bi-based materials. 
\end{enumerate}



Figure \ref{fig:groups}E shows the apical distance versus in-plane distance. The data points are coloured according to their Tc with red colours indicating higher Tc and blue colours indicating lower Tc.
It is clear that large areas of the apical/in-plane materials space, potentially yielding higher Tc materials, remain unexplored (This region is labelled `Unexplored Region' in Figure \ref{fig:groups}E and it occurs for high apical distance, i.e. above ~$2.5\angstrom$).

Past efforts have only been able to sample in small regions around known systems due to the limitations of perturbing systems with mechanical and chemical pressures.
A key limitation is that due to Poisson effects such methods influence both the a and c-lattice parameters preventing the exploration of trends in a one-factor-at-a-time manner.
Recently, new experiments have shown that it is possible to tune the a and c-lattice parameters independently allowing for unexplored regions of the materials space to be investigated \cite{choi_3d_2019, choi2020interfaces}; this vertically aligned nanocomposite (VAN) approach has led to enhanced Tcs of 50 K \cite{choi_3d_2019} and up to $\sim$120 K from magnetic measurements \cite{choi2020interfaces} in nano-engineered La\textsubscript{2}CuO\textsubscript{4-$\delta$} films relative to 40 K in the bulk (These points are highlighted in Figure \ref{fig:groups}E -- see Methods for how the apical distances for these samples were estimated).
These examples support the hypothesis that the unexplored region may yield higher Tc systems.

\section*{Discussions and Conclusion}

Having established the existence of a large unexplored region of the structure space of cuprate superconductors we believe that novel experimental approaches that allow for new regions of the apical/in-plane materials space to be probed would be fruitful to further understand structure-Tc trends in cuprates and potentially increase Tcs.

From the results presented here we believe that the 3D strain engineering of Bi2201 or Bi2212 systems are of particular interest because they lie at the base of the unexplored region with >$2.5\angstrom$ apical distances (Figure \ref{fig:groups}E).
3D strain engineering using VAN is suitable for Bi2201 and Bi2212 because they can be made in-situ via epitaxial growth methods \cite{zhu1993situ, nane2016effects}\textcolor{black}{, however, other methods such as non-linear phononics might also be appropriate \cite{mankowsky2014nonlinear}}.
Such experiments would allow greater insight into whether the high Tcs of Hg and Tl based cuprates are due to structural effects from the large apical distance or intrinsic electronic effects of the Hg and Tl cations.
\textcolor{black}{When attempting to optimise Tc, it should be noted that both Bi2201 and Bi2212 are known to benefit from substitutional doping \cite{eisaki2004effect}. This potential need for substitutional doping is important to consider in the selection of suitable materials for the substrate and matrix within the VAN setup. Bi2212 is also known to naturally exhibit crystal ``super-modulation'' which manifests as large variations in the Cu-O apical distance at the unit cell level \cite{slezak2008imaging}.}


Of interest also, is whether it would be possible to increase Tcs in Hg1212 or Hg1223 thin films in a manner that reduces the in-plane Cu-O distance whilst constraining the apical Cu-O distance to remain high.
This would allow us to move into Unexplored Region of Figure \ref{fig:groups}E from its right hand edge (from the cluster of red points at the highest apical distance values), rather than moving up from its bottom edge as proposed for Bi-based systems.
However, we believe this approach will be more challenging due to issues that arise when growing Hg-Ba-Ca-Cu-O thin films \cite{de2006effect}. 


Finally, we note that more systematic integration of existence data sources, deposition of new data and novel data mining efforts \cite{swain2016chemdataextractor} are desirable to improve superconductivity databases.
Consolidating the vast amount of information available in the literature into a comprehensive source, containing critical temperatures alongside atomic structures, would facilitate the improved application of materials informatics approaches.  \textcolor{black}{Critically, such resources would enable direct consideration of how  distributions of bond distances, and variations in bond angles e.g. buckling of the CuO\textsubscript{2} planes affect Tc.}

\section*{Methods}
\subsection{Examination of the variation of prototypical cuprates under strain}
Density functional theory calculations were used to model how cuprates might behave under strain.
Although standard DFT may not fully describe the electronic structure related to the superconducting states, we expect it still gives reasonable estimate for the strain responses.
The plane wave pseudopotential code \texttt{CASTEP} \cite{clark_first_2005} was used with the PBE exchange-correlation functional \cite{perdew_generalized_1996}. 
A plane wave cut off energy of 700 eV was used.
Monkhorst-Pack grids were used for sampling the reciprocal space with k-point spacing less than $2\pi \times 0.05 \angstrom^{-1}$.
On-the-fly generated core-corrected ultrasoft pseudopotentials \cite{vanderbilt_soft_1990} from \texttt{CASTEP}'s  \texttt{C18} library were used.
The equilibrium cell volumes of the structures were optimised with residual stresses less than 0.05 GPa. 
Once the equilibrium cell volumes were obtained, following optimisations were performed with fix cell sizes corresponding to strains in the c and a-b directions ranging from -4\% to +4\%. 
The ionic positions were relaxed until the maximum force was less than 0.01 eV$\angstrom^{-1}$ in all calculations.
The \texttt{AiiDA} framework was used to manage and automate the calculations \cite{pizzi_aiida_2016, huber_aiida_2020}.

\subsection{Estimation of the apical distance in VAN systems}
In \cite{choi_3d_2019} a Tc of 50K and a and c-lattice parameters of 3.79-3.76$\angstrom$ and 13.20-13.28$\angstrom$ are reported for the La\textsubscript{2}CuO\textsubscript{4-$\delta$}/LaCuO\textsubscript{3} (c-214/a-113) interfacial region.
The presence of domain matching in the structure suggests uniform stress along the c-axis, therefore, the apical distance can be estimated using the linear calibration model for the La214 (T) family. This gives an estimate of 2.40-2.43$\angstrom$ for the apical distance.

In \cite{choi2020interfaces} a weak magnetic signature for superconductivity at 120K is reported for a c-aligned La\textsubscript{2}CuO\textsubscript{4-$\delta$}/ a-aligned La\textsubscript{2}CuO\textsubscript{4-$\delta$} (c-214/a-214) interface.
Here there is La-block matching at the interface, rather than domain matching, suggesting a non-uniform stress.
The La-block is believed to be $4.00 \pm 0.01\angstrom$ at the interface - much larger than the average of 3.67$\angstrom$ for the ICSD La214 (T) reference data.
As this estimate requires a large degree of extrapolation we cannot justify a linear model.
Instead, we derive an estimate for the apical distance from considering the offset from the top of the La-block to the apical oxygen.
This offset is strongly peaked around 0.56$\angstrom$ giving an estimate of 2.56$\angstrom$ with a 90\% confidence interval of 2.51-2.62$\angstrom$.

\subsection*{Data and Code availability}
The processed data and the processing and plotting code used to analyse it are available from \href{www.github.com/comprhys/apical}{www.github.com/comprhys/apical}.

\section*{Acknowledgements}
R.E.A.G. and A.A.L. acknowledge the support of the Winton Programme for the Physics of Sustainability.
A.A.L. acknowledges support from the Royal Society. 
J.L.M-D. acknowledges funding from the Leverhulme Trust grant RPG-2020-041, U.S. Office of Naval Research Global grant N62909-18-1-2092, the Royal Academy of Engineering grant CIET1819\_24 and the Winton Foundation.
B.Z. acknowledges the support of the Cambridge Commonwealth, European and International Trust, the China Scholarship Council and funding from EPSRC grant EP/P007767/1 (CAM-IES).
This work was performed using resources provided by the Cambridge Service for Data Driven Discovery (CSD3) operated by the University of Cambridge Research Computing Service (www.csd3.cam.ac.uk), provided by Dell EMC and Intel using Tier-2 funding from the Engineering and Physical Sciences Research Council (capital grant EP/P020259/1), and DiRAC funding from the Science and Technology Facilities Council (www.dirac.ac.uk).

\bibliography{references.bib}

\begin{thebibliography}{43}%
\makeatletter
\providecommand \@ifxundefined [1]{%
 \@ifx{#1\undefined}
}%
\providecommand \@ifnum [1]{%
 \ifnum #1\expandafter \@firstoftwo
 \else \expandafter \@secondoftwo
 \fi
}%
\providecommand \@ifx [1]{%
 \ifx #1\expandafter \@firstoftwo
 \else \expandafter \@secondoftwo
 \fi
}%
\providecommand \natexlab [1]{#1}%
\providecommand \enquote  [1]{``#1''}%
\providecommand \bibnamefont  [1]{#1}%
\providecommand \bibfnamefont [1]{#1}%
\providecommand \citenamefont [1]{#1}%
\providecommand \href@noop [0]{\@secondoftwo}%
\providecommand \href [0]{\begingroup \@sanitize@url \@href}%
\providecommand \@href[1]{\@@startlink{#1}\@@href}%
\providecommand \@@href[1]{\endgroup#1\@@endlink}%
\providecommand \@sanitize@url [0]{\catcode `\\12\catcode `\$12\catcode
  `\&12\catcode `\#12\catcode `\^12\catcode `\_12\catcode `\%12\relax}%
\providecommand \@@startlink[1]{}%
\providecommand \@@endlink[0]{}%
\providecommand \url  [0]{\begingroup\@sanitize@url \@url }%
\providecommand \@url [1]{\endgroup\@href {#1}{\urlprefix }}%
\providecommand \urlprefix  [0]{URL }%
\providecommand \Eprint [0]{\href }%
\providecommand \doibase [0]{http://dx.doi.org/}%
\providecommand \selectlanguage [0]{\@gobble}%
\providecommand \bibinfo  [0]{\@secondoftwo}%
\providecommand \bibfield  [0]{\@secondoftwo}%
\providecommand \translation [1]{[#1]}%
\providecommand \BibitemOpen [0]{}%
\providecommand \bibitemStop [0]{}%
\providecommand \bibitemNoStop [0]{.\EOS\space}%
\providecommand \EOS [0]{\spacefactor3000\relax}%
\providecommand \BibitemShut  [1]{\csname bibitem#1\endcsname}%
\let\auto@bib@innerbib\@empty
\bibitem [{\citenamefont {Bednorz}\ and\ \citenamefont
  {M{\"u}ller}(1986)}]{bednorz1986possible}%
  \BibitemOpen
  \bibfield  {author} {\bibinfo {author} {\bibfnamefont {J.~G.}\ \bibnamefont
  {Bednorz}}\ and\ \bibinfo {author} {\bibfnamefont {K.~A.}\ \bibnamefont
  {M{\"u}ller}},\ }\href@noop {} {\bibfield  {journal} {\bibinfo  {journal}
  {Zeitschrift f{\"u}r Physik B Condensed Matter}\ }\textbf {\bibinfo {volume}
  {64}},\ \bibinfo {pages} {189} (\bibinfo {year} {1986})}\BibitemShut
  {NoStop}%
\bibitem [{\citenamefont {Homes}\ \emph {et~al.}(2004)\citenamefont {Homes},
  \citenamefont {Dordevic}, \citenamefont {Strongin}, \citenamefont {Bonn},
  \citenamefont {Liang}, \citenamefont {Hardy}, \citenamefont {Komiya},
  \citenamefont {Ando}, \citenamefont {Yu}, \citenamefont {Kaneko} \emph
  {et~al.}}]{homes2004universal}%
  \BibitemOpen
  \bibfield  {author} {\bibinfo {author} {\bibfnamefont {C.}~\bibnamefont
  {Homes}}, \bibinfo {author} {\bibfnamefont {S.}~\bibnamefont {Dordevic}},
  \bibinfo {author} {\bibfnamefont {M.}~\bibnamefont {Strongin}}, \bibinfo
  {author} {\bibfnamefont {D.}~\bibnamefont {Bonn}}, \bibinfo {author}
  {\bibfnamefont {R.}~\bibnamefont {Liang}}, \bibinfo {author} {\bibfnamefont
  {W.}~\bibnamefont {Hardy}}, \bibinfo {author} {\bibfnamefont
  {S.}~\bibnamefont {Komiya}}, \bibinfo {author} {\bibfnamefont
  {Y.}~\bibnamefont {Ando}}, \bibinfo {author} {\bibfnamefont {G.}~\bibnamefont
  {Yu}}, \bibinfo {author} {\bibfnamefont {N.}~\bibnamefont {Kaneko}},  \emph
  {et~al.},\ }\href@noop {} {\bibfield  {journal} {\bibinfo  {journal}
  {Nature}\ }\textbf {\bibinfo {volume} {430}},\ \bibinfo {pages} {539}
  (\bibinfo {year} {2004})}\BibitemShut {NoStop}%
\bibitem [{\citenamefont {Uemura}\ \emph {et~al.}(1989)\citenamefont {Uemura},
  \citenamefont {Luke}, \citenamefont {Sternlieb}, \citenamefont {Brewer},
  \citenamefont {Carolan}, \citenamefont {Hardy}, \citenamefont {Kadono},
  \citenamefont {Kempton}, \citenamefont {Kiefl}, \citenamefont {Kreitzman}
  \emph {et~al.}}]{uemura1989universal}%
  \BibitemOpen
  \bibfield  {author} {\bibinfo {author} {\bibfnamefont {Y.}~\bibnamefont
  {Uemura}}, \bibinfo {author} {\bibfnamefont {G.}~\bibnamefont {Luke}},
  \bibinfo {author} {\bibfnamefont {B.}~\bibnamefont {Sternlieb}}, \bibinfo
  {author} {\bibfnamefont {J.}~\bibnamefont {Brewer}}, \bibinfo {author}
  {\bibfnamefont {J.}~\bibnamefont {Carolan}}, \bibinfo {author} {\bibfnamefont
  {W.}~\bibnamefont {Hardy}}, \bibinfo {author} {\bibfnamefont
  {R.}~\bibnamefont {Kadono}}, \bibinfo {author} {\bibfnamefont
  {J.}~\bibnamefont {Kempton}}, \bibinfo {author} {\bibfnamefont
  {R.}~\bibnamefont {Kiefl}}, \bibinfo {author} {\bibfnamefont
  {S.}~\bibnamefont {Kreitzman}},  \emph {et~al.},\ }\href@noop {} {\bibfield
  {journal} {\bibinfo  {journal} {Physical review letters}\ }\textbf {\bibinfo
  {volume} {62}},\ \bibinfo {pages} {2317} (\bibinfo {year}
  {1989})}\BibitemShut {NoStop}%
\bibitem [{\citenamefont {Ohta}\ \emph {et~al.}(1991)\citenamefont {Ohta},
  \citenamefont {Tohyama},\ and\ \citenamefont {Maekawa}}]{ohta1991apex}%
  \BibitemOpen
  \bibfield  {author} {\bibinfo {author} {\bibfnamefont {Y.}~\bibnamefont
  {Ohta}}, \bibinfo {author} {\bibfnamefont {T.}~\bibnamefont {Tohyama}}, \
  and\ \bibinfo {author} {\bibfnamefont {S.}~\bibnamefont {Maekawa}},\
  }\href@noop {} {\bibfield  {journal} {\bibinfo  {journal} {Physical Review
  B}\ }\textbf {\bibinfo {volume} {43}},\ \bibinfo {pages} {2968} (\bibinfo
  {year} {1991})}\BibitemShut {NoStop}%
\bibitem [{\citenamefont {Rao}\ and\ \citenamefont
  {Ganguli}(1995)}]{rao_structureproperty_1995}%
  \BibitemOpen
  \bibfield  {author} {\bibinfo {author} {\bibfnamefont {C.~N.~R.}\
  \bibnamefont {Rao}}\ and\ \bibinfo {author} {\bibfnamefont {A.~K.}\
  \bibnamefont {Ganguli}},\ }\href {\doibase 10.1039/CS9952400001} {\bibfield
  {journal} {\bibinfo  {journal} {Chemical Society Reviews}\ }\textbf {\bibinfo
  {volume} {24}},\ \bibinfo {pages} {1} (\bibinfo {year} {1995})}\BibitemShut
  {NoStop}%
\bibitem [{\citenamefont {MacManus-Driscoll}\ and\ \citenamefont
  {Wimbush}(2011)}]{macmanus-driscoll_future_2011}%
  \BibitemOpen
  \bibfield  {author} {\bibinfo {author} {\bibfnamefont {J.~L.}\ \bibnamefont
  {MacManus-Driscoll}}\ and\ \bibinfo {author} {\bibfnamefont {S.~C.}\
  \bibnamefont {Wimbush}},\ }\href {\doibase 10.1109/TASC.2010.2100343}
  {\bibfield  {journal} {\bibinfo  {journal} {IEEE Transactions on Applied
  Superconductivity}\ }\textbf {\bibinfo {volume} {21}},\ \bibinfo {pages}
  {2495} (\bibinfo {year} {2011})}\BibitemShut {NoStop}%
\bibitem [{\citenamefont {Choi}\ \emph {et~al.}(2019)\citenamefont {Choi},
  \citenamefont {Bernardo}, \citenamefont {Zhu}, \citenamefont {Lu},
  \citenamefont {Alpern}, \citenamefont {Zhang}, \citenamefont {Shapira},
  \citenamefont {Feighan}, \citenamefont {Sun}, \citenamefont {Robinson},
  \citenamefont {Paltiel}, \citenamefont {Millo}, \citenamefont {Wang},
  \citenamefont {Jia},\ and\ \citenamefont {MacManus-Driscoll}}]{choi_3d_2019}%
  \BibitemOpen
  \bibfield  {author} {\bibinfo {author} {\bibfnamefont {E.-M.}\ \bibnamefont
  {Choi}}, \bibinfo {author} {\bibfnamefont {A.~D.}\ \bibnamefont {Bernardo}},
  \bibinfo {author} {\bibfnamefont {B.}~\bibnamefont {Zhu}}, \bibinfo {author}
  {\bibfnamefont {P.}~\bibnamefont {Lu}}, \bibinfo {author} {\bibfnamefont
  {H.}~\bibnamefont {Alpern}}, \bibinfo {author} {\bibfnamefont {K.~H.~L.}\
  \bibnamefont {Zhang}}, \bibinfo {author} {\bibfnamefont {T.}~\bibnamefont
  {Shapira}}, \bibinfo {author} {\bibfnamefont {J.}~\bibnamefont {Feighan}},
  \bibinfo {author} {\bibfnamefont {X.}~\bibnamefont {Sun}}, \bibinfo {author}
  {\bibfnamefont {J.}~\bibnamefont {Robinson}}, \bibinfo {author}
  {\bibfnamefont {Y.}~\bibnamefont {Paltiel}}, \bibinfo {author} {\bibfnamefont
  {O.}~\bibnamefont {Millo}}, \bibinfo {author} {\bibfnamefont
  {H.}~\bibnamefont {Wang}}, \bibinfo {author} {\bibfnamefont {Q.}~\bibnamefont
  {Jia}}, \ and\ \bibinfo {author} {\bibfnamefont {J.~L.}\ \bibnamefont
  {MacManus-Driscoll}},\ }\href {\doibase 10.1126/sciadv.aav5532} {\bibfield
  {journal} {\bibinfo  {journal} {Science Advances}\ }\textbf {\bibinfo
  {volume} {5}},\ \bibinfo {pages} {eaav5532} (\bibinfo {year}
  {2019})}\BibitemShut {NoStop}%
\bibitem [{\citenamefont {Choi}\ \emph {et~al.}(2020)\citenamefont {Choi},
  \citenamefont {Zhu}, \citenamefont {Lu}, \citenamefont {Feighan},
  \citenamefont {Sun}, \citenamefont {Wang},\ and\ \citenamefont
  {MacManus-Driscoll}}]{choi2020interfaces}%
  \BibitemOpen
  \bibfield  {author} {\bibinfo {author} {\bibfnamefont {E.-M.}\ \bibnamefont
  {Choi}}, \bibinfo {author} {\bibfnamefont {B.}~\bibnamefont {Zhu}}, \bibinfo
  {author} {\bibfnamefont {P.}~\bibnamefont {Lu}}, \bibinfo {author}
  {\bibfnamefont {J.}~\bibnamefont {Feighan}}, \bibinfo {author} {\bibfnamefont
  {X.}~\bibnamefont {Sun}}, \bibinfo {author} {\bibfnamefont {H.}~\bibnamefont
  {Wang}}, \ and\ \bibinfo {author} {\bibfnamefont {J.~L.}\ \bibnamefont
  {MacManus-Driscoll}},\ }\href {\doibase 10.1039/C9NR04996G} {\bibfield
  {journal} {\bibinfo  {journal} {Nanoscale}\ }\textbf {\bibinfo {volume}
  {12}},\ \bibinfo {pages} {3157} (\bibinfo {year} {2020})}\BibitemShut
  {NoStop}%
\bibitem [{\citenamefont {Pavarini}\ \emph {et~al.}(2001)\citenamefont
  {Pavarini}, \citenamefont {Dasgupta}, \citenamefont {Saha-Dasgupta},
  \citenamefont {Jepsen},\ and\ \citenamefont {Andersen}}]{pavarini2001band}%
  \BibitemOpen
  \bibfield  {author} {\bibinfo {author} {\bibfnamefont {E.}~\bibnamefont
  {Pavarini}}, \bibinfo {author} {\bibfnamefont {I.}~\bibnamefont {Dasgupta}},
  \bibinfo {author} {\bibfnamefont {T.}~\bibnamefont {Saha-Dasgupta}}, \bibinfo
  {author} {\bibfnamefont {O.}~\bibnamefont {Jepsen}}, \ and\ \bibinfo {author}
  {\bibfnamefont {O.}~\bibnamefont {Andersen}},\ }\href@noop {} {\bibfield
  {journal} {\bibinfo  {journal} {Physical review letters}\ }\textbf {\bibinfo
  {volume} {87}},\ \bibinfo {pages} {047003} (\bibinfo {year}
  {2001})}\BibitemShut {NoStop}%
\bibitem [{\citenamefont {Pop{\v{c}}evi{\'c}}\ \emph
  {et~al.}(2018)\citenamefont {Pop{\v{c}}evi{\'c}}, \citenamefont {Pelc},
  \citenamefont {Tang}, \citenamefont {Velebit}, \citenamefont {Anderson},
  \citenamefont {Nagarajan}, \citenamefont {Yu}, \citenamefont {Po{\v{z}}ek},
  \citenamefont {Bari{\v{s}}i{\'c}},\ and\ \citenamefont
  {Greven}}]{popvcevic2018percolative}%
  \BibitemOpen
  \bibfield  {author} {\bibinfo {author} {\bibfnamefont {P.}~\bibnamefont
  {Pop{\v{c}}evi{\'c}}}, \bibinfo {author} {\bibfnamefont {D.}~\bibnamefont
  {Pelc}}, \bibinfo {author} {\bibfnamefont {Y.}~\bibnamefont {Tang}}, \bibinfo
  {author} {\bibfnamefont {K.}~\bibnamefont {Velebit}}, \bibinfo {author}
  {\bibfnamefont {Z.}~\bibnamefont {Anderson}}, \bibinfo {author}
  {\bibfnamefont {V.}~\bibnamefont {Nagarajan}}, \bibinfo {author}
  {\bibfnamefont {G.}~\bibnamefont {Yu}}, \bibinfo {author} {\bibfnamefont
  {M.}~\bibnamefont {Po{\v{z}}ek}}, \bibinfo {author} {\bibfnamefont
  {N.}~\bibnamefont {Bari{\v{s}}i{\'c}}}, \ and\ \bibinfo {author}
  {\bibfnamefont {M.}~\bibnamefont {Greven}},\ }\href@noop {} {\bibfield
  {journal} {\bibinfo  {journal} {npj Quantum Materials}\ }\textbf {\bibinfo
  {volume} {3}},\ \bibinfo {pages} {1} (\bibinfo {year} {2018})}\BibitemShut
  {NoStop}%
\bibitem [{\citenamefont {Pelc}\ \emph {et~al.}(2018)\citenamefont {Pelc},
  \citenamefont {Vu{\v{c}}kovi{\'c}}, \citenamefont {Grbi{\'c}}, \citenamefont
  {Po{\v{z}}ek}, \citenamefont {Yu}, \citenamefont {Sasagawa}, \citenamefont
  {Greven},\ and\ \citenamefont {Bari{\v{s}}i{\'c}}}]{pelc2018emergence}%
  \BibitemOpen
  \bibfield  {author} {\bibinfo {author} {\bibfnamefont {D.}~\bibnamefont
  {Pelc}}, \bibinfo {author} {\bibfnamefont {M.}~\bibnamefont
  {Vu{\v{c}}kovi{\'c}}}, \bibinfo {author} {\bibfnamefont {M.~S.}\ \bibnamefont
  {Grbi{\'c}}}, \bibinfo {author} {\bibfnamefont {M.}~\bibnamefont
  {Po{\v{z}}ek}}, \bibinfo {author} {\bibfnamefont {G.}~\bibnamefont {Yu}},
  \bibinfo {author} {\bibfnamefont {T.}~\bibnamefont {Sasagawa}}, \bibinfo
  {author} {\bibfnamefont {M.}~\bibnamefont {Greven}}, \ and\ \bibinfo {author}
  {\bibfnamefont {N.}~\bibnamefont {Bari{\v{s}}i{\'c}}},\ }\href@noop {}
  {\bibfield  {journal} {\bibinfo  {journal} {Nature communications}\ }\textbf
  {\bibinfo {volume} {9}},\ \bibinfo {pages} {1} (\bibinfo {year}
  {2018})}\BibitemShut {NoStop}%
\bibitem [{\citenamefont {Pelc}\ \emph {et~al.}(2019)\citenamefont {Pelc},
  \citenamefont {Anderson}, \citenamefont {Yu}, \citenamefont {Leighton},\ and\
  \citenamefont {Greven}}]{pelc2019universal}%
  \BibitemOpen
  \bibfield  {author} {\bibinfo {author} {\bibfnamefont {D.}~\bibnamefont
  {Pelc}}, \bibinfo {author} {\bibfnamefont {Z.}~\bibnamefont {Anderson}},
  \bibinfo {author} {\bibfnamefont {B.}~\bibnamefont {Yu}}, \bibinfo {author}
  {\bibfnamefont {C.}~\bibnamefont {Leighton}}, \ and\ \bibinfo {author}
  {\bibfnamefont {M.}~\bibnamefont {Greven}},\ }\href@noop {} {\bibfield
  {journal} {\bibinfo  {journal} {Nature communications}\ }\textbf {\bibinfo
  {volume} {10}},\ \bibinfo {pages} {1} (\bibinfo {year} {2019})}\BibitemShut
  {NoStop}%
\bibitem [{\citenamefont {Plakida}(2010)}]{plakida2010high}%
  \BibitemOpen
  \bibfield  {author} {\bibinfo {author} {\bibfnamefont {N.}~\bibnamefont
  {Plakida}},\ }\href@noop {} {\emph {\bibinfo {title} {High-Temperature
  Cuprate Superconductors: Experiment, Theory, and Applications}}},\ Vol.\
  \bibinfo {volume} {166}\ (\bibinfo  {publisher} {Springer Science \& Business
  Media},\ \bibinfo {year} {2010})\BibitemShut {NoStop}%
\bibitem [{\citenamefont {Fradkin}\ \emph {et~al.}(2015)\citenamefont
  {Fradkin}, \citenamefont {Kivelson},\ and\ \citenamefont
  {Tranquada}}]{fradkin2015colloquium}%
  \BibitemOpen
  \bibfield  {author} {\bibinfo {author} {\bibfnamefont {E.}~\bibnamefont
  {Fradkin}}, \bibinfo {author} {\bibfnamefont {S.~A.}\ \bibnamefont
  {Kivelson}}, \ and\ \bibinfo {author} {\bibfnamefont {J.~M.}\ \bibnamefont
  {Tranquada}},\ }\href@noop {} {\bibfield  {journal} {\bibinfo  {journal}
  {Reviews of Modern Physics}\ }\textbf {\bibinfo {volume} {87}},\ \bibinfo
  {pages} {457} (\bibinfo {year} {2015})}\BibitemShut {NoStop}%
\bibitem [{\citenamefont {Stanev}\ \emph {et~al.}(2018)\citenamefont {Stanev},
  \citenamefont {Oses}, \citenamefont {Kusne}, \citenamefont {Rodriguez},
  \citenamefont {Paglione}, \citenamefont {Curtarolo},\ and\ \citenamefont
  {Takeuchi}}]{stanev2018machine}%
  \BibitemOpen
  \bibfield  {author} {\bibinfo {author} {\bibfnamefont {V.}~\bibnamefont
  {Stanev}}, \bibinfo {author} {\bibfnamefont {C.}~\bibnamefont {Oses}},
  \bibinfo {author} {\bibfnamefont {A.~G.}\ \bibnamefont {Kusne}}, \bibinfo
  {author} {\bibfnamefont {E.}~\bibnamefont {Rodriguez}}, \bibinfo {author}
  {\bibfnamefont {J.}~\bibnamefont {Paglione}}, \bibinfo {author}
  {\bibfnamefont {S.}~\bibnamefont {Curtarolo}}, \ and\ \bibinfo {author}
  {\bibfnamefont {I.}~\bibnamefont {Takeuchi}},\ }\href@noop {} {\bibfield
  {journal} {\bibinfo  {journal} {npj Computational Materials}\ }\textbf
  {\bibinfo {volume} {4}},\ \bibinfo {pages} {29} (\bibinfo {year}
  {2018})}\BibitemShut {NoStop}%
\bibitem [{\citenamefont {Konno}\ \emph {et~al.}(2018)\citenamefont {Konno},
  \citenamefont {Kurokawa}, \citenamefont {Nabeshima}, \citenamefont
  {Sakishita}, \citenamefont {Ogawa}, \citenamefont {Hosako},\ and\
  \citenamefont {Maeda}}]{konno2018deep}%
  \BibitemOpen
  \bibfield  {author} {\bibinfo {author} {\bibfnamefont {T.}~\bibnamefont
  {Konno}}, \bibinfo {author} {\bibfnamefont {H.}~\bibnamefont {Kurokawa}},
  \bibinfo {author} {\bibfnamefont {F.}~\bibnamefont {Nabeshima}}, \bibinfo
  {author} {\bibfnamefont {Y.}~\bibnamefont {Sakishita}}, \bibinfo {author}
  {\bibfnamefont {R.}~\bibnamefont {Ogawa}}, \bibinfo {author} {\bibfnamefont
  {I.}~\bibnamefont {Hosako}}, \ and\ \bibinfo {author} {\bibfnamefont
  {A.}~\bibnamefont {Maeda}},\ }\href@noop {} {\enquote {\bibinfo {title} {Deep
  learning model for finding new superconductors},}\ } (\bibinfo {year}
  {2018}),\ \Eprint {http://arxiv.org/abs/1812.01995} {arXiv:1812.01995
  [cs.LG]} \BibitemShut {NoStop}%
\bibitem [{\citenamefont {Liu}\ \emph {et~al.}(2020)\citenamefont {Liu},
  \citenamefont {Kang}, \citenamefont {Zhu}, \citenamefont {Liu},\ and\
  \citenamefont {Guo}}]{liu2020material}%
  \BibitemOpen
  \bibfield  {author} {\bibinfo {author} {\bibfnamefont {Z.-L.}\ \bibnamefont
  {Liu}}, \bibinfo {author} {\bibfnamefont {P.}~\bibnamefont {Kang}}, \bibinfo
  {author} {\bibfnamefont {Y.}~\bibnamefont {Zhu}}, \bibinfo {author}
  {\bibfnamefont {L.}~\bibnamefont {Liu}}, \ and\ \bibinfo {author}
  {\bibfnamefont {H.}~\bibnamefont {Guo}},\ }\href@noop {} {\bibfield
  {journal} {\bibinfo  {journal} {APL Materials}\ }\textbf {\bibinfo {volume}
  {8}},\ \bibinfo {pages} {061104} (\bibinfo {year} {2020})}\BibitemShut
  {NoStop}%
\bibitem [{\citenamefont {Meredig}\ \emph {et~al.}(2018)\citenamefont
  {Meredig}, \citenamefont {Antono}, \citenamefont {Church}, \citenamefont
  {Hutchinson}, \citenamefont {Ling}, \citenamefont {Paradiso}, \citenamefont
  {Blaiszik}, \citenamefont {Foster}, \citenamefont {Gibbons}, \citenamefont
  {Hattrick-Simpers}, \citenamefont {Mehta},\ and\ \citenamefont
  {Ward}}]{meredig2018can}%
  \BibitemOpen
  \bibfield  {author} {\bibinfo {author} {\bibfnamefont {B.}~\bibnamefont
  {Meredig}}, \bibinfo {author} {\bibfnamefont {E.}~\bibnamefont {Antono}},
  \bibinfo {author} {\bibfnamefont {C.}~\bibnamefont {Church}}, \bibinfo
  {author} {\bibfnamefont {M.}~\bibnamefont {Hutchinson}}, \bibinfo {author}
  {\bibfnamefont {J.}~\bibnamefont {Ling}}, \bibinfo {author} {\bibfnamefont
  {S.}~\bibnamefont {Paradiso}}, \bibinfo {author} {\bibfnamefont
  {B.}~\bibnamefont {Blaiszik}}, \bibinfo {author} {\bibfnamefont
  {I.}~\bibnamefont {Foster}}, \bibinfo {author} {\bibfnamefont
  {B.}~\bibnamefont {Gibbons}}, \bibinfo {author} {\bibfnamefont
  {J.}~\bibnamefont {Hattrick-Simpers}}, \bibinfo {author} {\bibfnamefont
  {A.}~\bibnamefont {Mehta}}, \ and\ \bibinfo {author} {\bibfnamefont
  {L.}~\bibnamefont {Ward}},\ }\href {\doibase 10.1039/C8ME00012C} {\bibfield
  {journal} {\bibinfo  {journal} {Mol. Syst. Des. Eng.}\ }\textbf {\bibinfo
  {volume} {3}},\ \bibinfo {pages} {819} (\bibinfo {year} {2018})}\BibitemShut
  {NoStop}%
\bibitem [{\citenamefont {Kim}\ \emph {et~al.}(2018)\citenamefont {Kim},
  \citenamefont {Chen}, \citenamefont {Fitzhugh},\ and\ \citenamefont
  {Li}}]{kim2018apical}%
  \BibitemOpen
  \bibfield  {author} {\bibinfo {author} {\bibfnamefont {S.}~\bibnamefont
  {Kim}}, \bibinfo {author} {\bibfnamefont {X.}~\bibnamefont {Chen}}, \bibinfo
  {author} {\bibfnamefont {W.}~\bibnamefont {Fitzhugh}}, \ and\ \bibinfo
  {author} {\bibfnamefont {X.}~\bibnamefont {Li}},\ }\href@noop {} {\bibfield
  {journal} {\bibinfo  {journal} {Physical review letters}\ }\textbf {\bibinfo
  {volume} {121}},\ \bibinfo {pages} {157001} (\bibinfo {year}
  {2018})}\BibitemShut {NoStop}%
\bibitem [{\citenamefont {Hellenbrandt}(2004)}]{hellenbrandt2004inorganic}%
  \BibitemOpen
  \bibfield  {author} {\bibinfo {author} {\bibfnamefont {M.}~\bibnamefont
  {Hellenbrandt}},\ }\href@noop {} {\bibfield  {journal} {\bibinfo  {journal}
  {Crystallography Reviews}\ }\textbf {\bibinfo {volume} {10}},\ \bibinfo
  {pages} {17} (\bibinfo {year} {2004})}\BibitemShut {NoStop}%
\bibitem [{\citenamefont {Poole}\ \emph {et~al.}(2014)\citenamefont {Poole},
  \citenamefont {Prozorov}, \citenamefont {Farach},\ and\ \citenamefont
  {Creswick}}]{poole2014super}%
  \BibitemOpen
  \bibfield  {author} {\bibinfo {author} {\bibfnamefont {C.~P.}\ \bibnamefont
  {Poole}}, \bibinfo {author} {\bibfnamefont {R.}~\bibnamefont {Prozorov}},
  \bibinfo {author} {\bibfnamefont {H.~A.}\ \bibnamefont {Farach}}, \ and\
  \bibinfo {author} {\bibfnamefont {R.~J.}\ \bibnamefont {Creswick}},\
  }\href@noop {} {\emph {\bibinfo {title} {Superconductivity}}}\ (\bibinfo
  {publisher} {Elsevier},\ \bibinfo {year} {2014})\ pp.\ \bibinfo {pages}
  {677--726}\BibitemShut {NoStop}%
\bibitem [{\citenamefont {Eisaki}\ \emph {et~al.}(2004)\citenamefont {Eisaki},
  \citenamefont {Kaneko}, \citenamefont {Feng}, \citenamefont {Damascelli},
  \citenamefont {Mang}, \citenamefont {Shen}, \citenamefont {Shen},\ and\
  \citenamefont {Greven}}]{eisaki2004effect}%
  \BibitemOpen
  \bibfield  {author} {\bibinfo {author} {\bibfnamefont {H.}~\bibnamefont
  {Eisaki}}, \bibinfo {author} {\bibfnamefont {N.}~\bibnamefont {Kaneko}},
  \bibinfo {author} {\bibfnamefont {D.}~\bibnamefont {Feng}}, \bibinfo {author}
  {\bibfnamefont {A.}~\bibnamefont {Damascelli}}, \bibinfo {author}
  {\bibfnamefont {P.}~\bibnamefont {Mang}}, \bibinfo {author} {\bibfnamefont
  {K.}~\bibnamefont {Shen}}, \bibinfo {author} {\bibfnamefont {Z.-X.}\
  \bibnamefont {Shen}}, \ and\ \bibinfo {author} {\bibfnamefont
  {M.}~\bibnamefont {Greven}},\ }\href@noop {} {\bibfield  {journal} {\bibinfo
  {journal} {Physical Review B}\ }\textbf {\bibinfo {volume} {69}},\ \bibinfo
  {pages} {064512} (\bibinfo {year} {2004})}\BibitemShut {NoStop}%
\bibitem [{\citenamefont {Hyatt}\ \emph {et~al.}(1998)\citenamefont {Hyatt},
  \citenamefont {Jones}, \citenamefont {Gameson}, \citenamefont {Slaski},
  \citenamefont {Peacock}, \citenamefont {Hriljac},\ and\ \citenamefont
  {Edwards}}]{hyatt1998synthesis}%
  \BibitemOpen
  \bibfield  {author} {\bibinfo {author} {\bibfnamefont {N.~C.}\ \bibnamefont
  {Hyatt}}, \bibinfo {author} {\bibfnamefont {M.}~\bibnamefont {Jones}},
  \bibinfo {author} {\bibfnamefont {I.}~\bibnamefont {Gameson}}, \bibinfo
  {author} {\bibfnamefont {M.}~\bibnamefont {Slaski}}, \bibinfo {author}
  {\bibfnamefont {G.}~\bibnamefont {Peacock}}, \bibinfo {author} {\bibfnamefont
  {J.}~\bibnamefont {Hriljac}}, \ and\ \bibinfo {author} {\bibfnamefont
  {P.~P.}\ \bibnamefont {Edwards}},\ }\href@noop {} {\bibfield  {journal}
  {\bibinfo  {journal} {Journal of superconductivity}\ }\textbf {\bibinfo
  {volume} {11}},\ \bibinfo {pages} {141} (\bibinfo {year} {1998})}\BibitemShut
  {NoStop}%
\bibitem [{\citenamefont {Huber}(1992)}]{huber1992robust}%
  \BibitemOpen
  \bibfield  {author} {\bibinfo {author} {\bibfnamefont {P.~J.}\ \bibnamefont
  {Huber}},\ }in\ \href@noop {} {\emph {\bibinfo {booktitle} {Breakthroughs in
  statistics}}}\ (\bibinfo  {publisher} {Springer},\ \bibinfo {year} {1992})\
  pp.\ \bibinfo {pages} {492--518}\BibitemShut {NoStop}%
\bibitem [{\citenamefont {Lloyd}(1982)}]{lloyd1982least}%
  \BibitemOpen
  \bibfield  {author} {\bibinfo {author} {\bibfnamefont {S.}~\bibnamefont
  {Lloyd}},\ }\href@noop {} {\bibfield  {journal} {\bibinfo  {journal} {IEEE
  transactions on information theory}\ }\textbf {\bibinfo {volume} {28}},\
  \bibinfo {pages} {129} (\bibinfo {year} {1982})}\BibitemShut {NoStop}%
\bibitem [{\citenamefont {Di~Castro}\ and\ \citenamefont
  {Balestrino}(2018)}]{di2018superconductivity}%
  \BibitemOpen
  \bibfield  {author} {\bibinfo {author} {\bibfnamefont {D.}~\bibnamefont
  {Di~Castro}}\ and\ \bibinfo {author} {\bibfnamefont {G.}~\bibnamefont
  {Balestrino}},\ }\href@noop {} {\bibfield  {journal} {\bibinfo  {journal}
  {Superconductor Science and Technology}\ }\textbf {\bibinfo {volume} {31}},\
  \bibinfo {pages} {073001} (\bibinfo {year} {2018})}\BibitemShut {NoStop}%
\bibitem [{\citenamefont {Gao}\ \emph {et~al.}(1994)\citenamefont {Gao},
  \citenamefont {Xue}, \citenamefont {Chen}, \citenamefont {Xiong},
  \citenamefont {Meng}, \citenamefont {Ramirez}, \citenamefont {Chu},
  \citenamefont {Eggert},\ and\ \citenamefont
  {Mao}}]{gao1994superconductivity}%
  \BibitemOpen
  \bibfield  {author} {\bibinfo {author} {\bibfnamefont {L.}~\bibnamefont
  {Gao}}, \bibinfo {author} {\bibfnamefont {Y.}~\bibnamefont {Xue}}, \bibinfo
  {author} {\bibfnamefont {F.}~\bibnamefont {Chen}}, \bibinfo {author}
  {\bibfnamefont {Q.}~\bibnamefont {Xiong}}, \bibinfo {author} {\bibfnamefont
  {R.}~\bibnamefont {Meng}}, \bibinfo {author} {\bibfnamefont {D.}~\bibnamefont
  {Ramirez}}, \bibinfo {author} {\bibfnamefont {C.}~\bibnamefont {Chu}},
  \bibinfo {author} {\bibfnamefont {J.}~\bibnamefont {Eggert}}, \ and\ \bibinfo
  {author} {\bibfnamefont {H.}~\bibnamefont {Mao}},\ }\href@noop {} {\bibfield
  {journal} {\bibinfo  {journal} {Physical Review B}\ }\textbf {\bibinfo
  {volume} {50}},\ \bibinfo {pages} {4260} (\bibinfo {year}
  {1994})}\BibitemShut {NoStop}%
\bibitem [{\citenamefont {Yamamoto}\ \emph {et~al.}(2015)\citenamefont
  {Yamamoto}, \citenamefont {Takeshita}, \citenamefont {Terakura},\ and\
  \citenamefont {Tokura}}]{yamamoto2015high}%
  \BibitemOpen
  \bibfield  {author} {\bibinfo {author} {\bibfnamefont {A.}~\bibnamefont
  {Yamamoto}}, \bibinfo {author} {\bibfnamefont {N.}~\bibnamefont {Takeshita}},
  \bibinfo {author} {\bibfnamefont {C.}~\bibnamefont {Terakura}}, \ and\
  \bibinfo {author} {\bibfnamefont {Y.}~\bibnamefont {Tokura}},\ }\href@noop {}
  {\bibfield  {journal} {\bibinfo  {journal} {Nature communications}\ }\textbf
  {\bibinfo {volume} {6}},\ \bibinfo {pages} {1} (\bibinfo {year}
  {2015})}\BibitemShut {NoStop}%
\bibitem [{\citenamefont {Armstrong}\ \emph {et~al.}(1995)\citenamefont
  {Armstrong}, \citenamefont {David}, \citenamefont {Gameson}, \citenamefont
  {Edwards}, \citenamefont {Capponi}, \citenamefont {Bordet},\ and\
  \citenamefont {Marezio}}]{armstrong1995crystal}%
  \BibitemOpen
  \bibfield  {author} {\bibinfo {author} {\bibfnamefont {A.}~\bibnamefont
  {Armstrong}}, \bibinfo {author} {\bibfnamefont {W.}~\bibnamefont {David}},
  \bibinfo {author} {\bibfnamefont {I.}~\bibnamefont {Gameson}}, \bibinfo
  {author} {\bibfnamefont {P.}~\bibnamefont {Edwards}}, \bibinfo {author}
  {\bibfnamefont {J.}~\bibnamefont {Capponi}}, \bibinfo {author} {\bibfnamefont
  {P.}~\bibnamefont {Bordet}}, \ and\ \bibinfo {author} {\bibfnamefont
  {M.}~\bibnamefont {Marezio}},\ }\href@noop {} {\bibfield  {journal} {\bibinfo
   {journal} {Physical Review B}\ }\textbf {\bibinfo {volume} {52}},\ \bibinfo
  {pages} {15551} (\bibinfo {year} {1995})}\BibitemShut {NoStop}%
\bibitem [{\citenamefont {Gatt}\ \emph {et~al.}(1998)\citenamefont {Gatt},
  \citenamefont {Olsen}, \citenamefont {Gerward}, \citenamefont {Bryntse},
  \citenamefont {Kareiva}, \citenamefont {Panas},\ and\ \citenamefont
  {Johansson}}]{gatt1998pressure}%
  \BibitemOpen
  \bibfield  {author} {\bibinfo {author} {\bibfnamefont {R.}~\bibnamefont
  {Gatt}}, \bibinfo {author} {\bibfnamefont {J.}~\bibnamefont {Olsen}},
  \bibinfo {author} {\bibfnamefont {L.}~\bibnamefont {Gerward}}, \bibinfo
  {author} {\bibfnamefont {I.}~\bibnamefont {Bryntse}}, \bibinfo {author}
  {\bibfnamefont {A.}~\bibnamefont {Kareiva}}, \bibinfo {author} {\bibfnamefont
  {I.}~\bibnamefont {Panas}}, \ and\ \bibinfo {author} {\bibfnamefont {L.-G.}\
  \bibnamefont {Johansson}},\ }\href@noop {} {\bibfield  {journal} {\bibinfo
  {journal} {Physical Review B}\ }\textbf {\bibinfo {volume} {57}},\ \bibinfo
  {pages} {13922} (\bibinfo {year} {1998})}\BibitemShut {NoStop}%
\bibitem [{\citenamefont {Volkova}\ \emph {et~al.}(2000)\citenamefont
  {Volkova}, \citenamefont {Polishchuk}, \citenamefont {Magarill},\ and\
  \citenamefont {Sobolev}}]{volkova2000correlation}%
  \BibitemOpen
  \bibfield  {author} {\bibinfo {author} {\bibfnamefont {L.}~\bibnamefont
  {Volkova}}, \bibinfo {author} {\bibfnamefont {S.}~\bibnamefont {Polishchuk}},
  \bibinfo {author} {\bibfnamefont {S.}~\bibnamefont {Magarill}}, \ and\
  \bibinfo {author} {\bibfnamefont {A.}~\bibnamefont {Sobolev}},\ }\href@noop
  {} {\bibfield  {journal} {\bibinfo  {journal} {Inorganic materials}\ }\textbf
  {\bibinfo {volume} {36}},\ \bibinfo {pages} {919} (\bibinfo {year}
  {2000})}\BibitemShut {NoStop}%
\bibitem [{\citenamefont {Yoshizaki}\ \emph {et~al.}(2007)\citenamefont
  {Yoshizaki}, \citenamefont {Nakajima},\ and\ \citenamefont
  {Tange}}]{yoshizaki2007substitution}%
  \BibitemOpen
  \bibfield  {author} {\bibinfo {author} {\bibfnamefont {R.}~\bibnamefont
  {Yoshizaki}}, \bibinfo {author} {\bibfnamefont {T.}~\bibnamefont {Nakajima}},
  \ and\ \bibinfo {author} {\bibfnamefont {M.}~\bibnamefont {Tange}},\
  }\href@noop {} {\bibfield  {journal} {\bibinfo  {journal} {Japanese journal
  of applied physics}\ }\textbf {\bibinfo {volume} {46}},\ \bibinfo {pages}
  {L167} (\bibinfo {year} {2007})}\BibitemShut {NoStop}%
\bibitem [{\citenamefont {Zhu}\ \emph {et~al.}(1993)\citenamefont {Zhu},
  \citenamefont {Lowndes}, \citenamefont {Chakoumakos}, \citenamefont {Budai},
  \citenamefont {Christen}, \citenamefont {Zheng}, \citenamefont {Jones},\ and\
  \citenamefont {Warmack}}]{zhu1993situ}%
  \BibitemOpen
  \bibfield  {author} {\bibinfo {author} {\bibfnamefont {S.}~\bibnamefont
  {Zhu}}, \bibinfo {author} {\bibfnamefont {D.}~\bibnamefont {Lowndes}},
  \bibinfo {author} {\bibfnamefont {B.}~\bibnamefont {Chakoumakos}}, \bibinfo
  {author} {\bibfnamefont {J.}~\bibnamefont {Budai}}, \bibinfo {author}
  {\bibfnamefont {D.}~\bibnamefont {Christen}}, \bibinfo {author}
  {\bibfnamefont {X.-Y.}\ \bibnamefont {Zheng}}, \bibinfo {author}
  {\bibfnamefont {E.}~\bibnamefont {Jones}}, \ and\ \bibinfo {author}
  {\bibfnamefont {B.}~\bibnamefont {Warmack}},\ }\href@noop {} {\bibfield
  {journal} {\bibinfo  {journal} {Applied physics letters}\ }\textbf {\bibinfo
  {volume} {63}},\ \bibinfo {pages} {409} (\bibinfo {year} {1993})}\BibitemShut
  {NoStop}%
\bibitem [{\citenamefont {Nane}\ \emph {et~al.}(2016)\citenamefont {Nane},
  \citenamefont {{\"O}z{\c{c}}elik},\ and\ \citenamefont
  {Abukay}}]{nane2016effects}%
  \BibitemOpen
  \bibfield  {author} {\bibinfo {author} {\bibfnamefont {O.}~\bibnamefont
  {Nane}}, \bibinfo {author} {\bibfnamefont {B.}~\bibnamefont
  {{\"O}z{\c{c}}elik}}, \ and\ \bibinfo {author} {\bibfnamefont
  {D.}~\bibnamefont {Abukay}},\ }\href@noop {} {\bibfield  {journal} {\bibinfo
  {journal} {Ceramics International}\ }\textbf {\bibinfo {volume} {42}},\
  \bibinfo {pages} {5778} (\bibinfo {year} {2016})}\BibitemShut {NoStop}%
\bibitem [{\citenamefont {Mankowsky}\ \emph {et~al.}(2014)\citenamefont
  {Mankowsky}, \citenamefont {Subedi}, \citenamefont {F{\"o}rst}, \citenamefont
  {Mariager}, \citenamefont {Chollet}, \citenamefont {Lemke}, \citenamefont
  {Robinson}, \citenamefont {Glownia}, \citenamefont {Minitti}, \citenamefont
  {Frano} \emph {et~al.}}]{mankowsky2014nonlinear}%
  \BibitemOpen
  \bibfield  {author} {\bibinfo {author} {\bibfnamefont {R.}~\bibnamefont
  {Mankowsky}}, \bibinfo {author} {\bibfnamefont {A.}~\bibnamefont {Subedi}},
  \bibinfo {author} {\bibfnamefont {M.}~\bibnamefont {F{\"o}rst}}, \bibinfo
  {author} {\bibfnamefont {S.~O.}\ \bibnamefont {Mariager}}, \bibinfo {author}
  {\bibfnamefont {M.}~\bibnamefont {Chollet}}, \bibinfo {author} {\bibfnamefont
  {H.}~\bibnamefont {Lemke}}, \bibinfo {author} {\bibfnamefont {J.~S.}\
  \bibnamefont {Robinson}}, \bibinfo {author} {\bibfnamefont {J.~M.}\
  \bibnamefont {Glownia}}, \bibinfo {author} {\bibfnamefont {M.~P.}\
  \bibnamefont {Minitti}}, \bibinfo {author} {\bibfnamefont {A.}~\bibnamefont
  {Frano}},  \emph {et~al.},\ }\href@noop {} {\bibfield  {journal} {\bibinfo
  {journal} {Nature}\ }\textbf {\bibinfo {volume} {516}},\ \bibinfo {pages}
  {71} (\bibinfo {year} {2014})}\BibitemShut {NoStop}%
\bibitem [{\citenamefont {Slezak}\ \emph {et~al.}(2008)\citenamefont {Slezak},
  \citenamefont {Lee}, \citenamefont {Wang}, \citenamefont {McElroy},
  \citenamefont {Fujita}, \citenamefont {Andersen}, \citenamefont {Hirschfeld},
  \citenamefont {Eisaki}, \citenamefont {Uchida},\ and\ \citenamefont
  {Davis}}]{slezak2008imaging}%
  \BibitemOpen
  \bibfield  {author} {\bibinfo {author} {\bibfnamefont {J.}~\bibnamefont
  {Slezak}}, \bibinfo {author} {\bibfnamefont {J.}~\bibnamefont {Lee}},
  \bibinfo {author} {\bibfnamefont {M.}~\bibnamefont {Wang}}, \bibinfo {author}
  {\bibfnamefont {K.}~\bibnamefont {McElroy}}, \bibinfo {author} {\bibfnamefont
  {K.}~\bibnamefont {Fujita}}, \bibinfo {author} {\bibfnamefont
  {B.}~\bibnamefont {Andersen}}, \bibinfo {author} {\bibfnamefont
  {P.}~\bibnamefont {Hirschfeld}}, \bibinfo {author} {\bibfnamefont
  {H.}~\bibnamefont {Eisaki}}, \bibinfo {author} {\bibfnamefont
  {S.}~\bibnamefont {Uchida}}, \ and\ \bibinfo {author} {\bibfnamefont
  {J.}~\bibnamefont {Davis}},\ }\href@noop {} {\bibfield  {journal} {\bibinfo
  {journal} {Proceedings of the National Academy of Sciences}\ }\textbf
  {\bibinfo {volume} {105}},\ \bibinfo {pages} {3203} (\bibinfo {year}
  {2008})}\BibitemShut {NoStop}%
\bibitem [{\citenamefont {De~Barros}\ \emph {et~al.}(2006)\citenamefont
  {De~Barros}, \citenamefont {Ortega}, \citenamefont {Peroz}, \citenamefont
  {Weiss},\ and\ \citenamefont {Odier}}]{de2006effect}%
  \BibitemOpen
  \bibfield  {author} {\bibinfo {author} {\bibfnamefont {D.}~\bibnamefont
  {De~Barros}}, \bibinfo {author} {\bibfnamefont {L.}~\bibnamefont {Ortega}},
  \bibinfo {author} {\bibfnamefont {C.}~\bibnamefont {Peroz}}, \bibinfo
  {author} {\bibfnamefont {F.}~\bibnamefont {Weiss}}, \ and\ \bibinfo {author}
  {\bibfnamefont {P.}~\bibnamefont {Odier}},\ }\href@noop {} {\bibfield
  {journal} {\bibinfo  {journal} {Physica C: Superconductivity}\ }\textbf
  {\bibinfo {volume} {440}},\ \bibinfo {pages} {45} (\bibinfo {year}
  {2006})}\BibitemShut {NoStop}%
\bibitem [{\citenamefont {Swain}\ and\ \citenamefont
  {Cole}(2016)}]{swain2016chemdataextractor}%
  \BibitemOpen
  \bibfield  {author} {\bibinfo {author} {\bibfnamefont {M.~C.}\ \bibnamefont
  {Swain}}\ and\ \bibinfo {author} {\bibfnamefont {J.~M.}\ \bibnamefont
  {Cole}},\ }\href@noop {} {\bibfield  {journal} {\bibinfo  {journal} {Journal
  of chemical information and modeling}\ }\textbf {\bibinfo {volume} {56}},\
  \bibinfo {pages} {1894} (\bibinfo {year} {2016})}\BibitemShut {NoStop}%
\bibitem [{\citenamefont {Clark}\ \emph {et~al.}(2005)\citenamefont {Clark},
  \citenamefont {Segall}, \citenamefont {Pickard}, \citenamefont {Hasnip},
  \citenamefont {Probert}, \citenamefont {Refson},\ and\ \citenamefont
  {Payne}}]{clark_first_2005}%
  \BibitemOpen
  \bibfield  {author} {\bibinfo {author} {\bibfnamefont {S.~J.}\ \bibnamefont
  {Clark}}, \bibinfo {author} {\bibfnamefont {M.~D.}\ \bibnamefont {Segall}},
  \bibinfo {author} {\bibfnamefont {C.~J.}\ \bibnamefont {Pickard}}, \bibinfo
  {author} {\bibfnamefont {P.~J.}\ \bibnamefont {Hasnip}}, \bibinfo {author}
  {\bibfnamefont {M.~I.~J.}\ \bibnamefont {Probert}}, \bibinfo {author}
  {\bibfnamefont {K.}~\bibnamefont {Refson}}, \ and\ \bibinfo {author}
  {\bibfnamefont {M.~C.}\ \bibnamefont {Payne}},\ }\href {\doibase
  10.1524/zkri.220.5.567.65075} {\bibfield  {journal} {\bibinfo  {journal}
  {Zeitschrift f{\"u}r Kristallographie}\ }\textbf {\bibinfo {volume} {220}},\
  \bibinfo {pages} {567} (\bibinfo {year} {2005})}\BibitemShut {NoStop}%
\bibitem [{\citenamefont {Perdew}\ \emph {et~al.}(1996)\citenamefont {Perdew},
  \citenamefont {Burke},\ and\ \citenamefont
  {Ernzerhof}}]{perdew_generalized_1996}%
  \BibitemOpen
  \bibfield  {author} {\bibinfo {author} {\bibfnamefont {J.~P.}\ \bibnamefont
  {Perdew}}, \bibinfo {author} {\bibfnamefont {K.}~\bibnamefont {Burke}}, \
  and\ \bibinfo {author} {\bibfnamefont {M.}~\bibnamefont {Ernzerhof}},\ }\href
  {\doibase 10.1103/PhysRevLett.77.3865} {\bibfield  {journal} {\bibinfo
  {journal} {Physical Review Letters}\ }\textbf {\bibinfo {volume} {77}},\
  \bibinfo {pages} {3865} (\bibinfo {year} {1996})}\BibitemShut {NoStop}%
\bibitem [{\citenamefont {Vanderbilt}(1990)}]{vanderbilt_soft_1990}%
  \BibitemOpen
  \bibfield  {author} {\bibinfo {author} {\bibfnamefont {D.}~\bibnamefont
  {Vanderbilt}},\ }\href {\doibase 10.1103/PhysRevB.41.7892} {\bibfield
  {journal} {\bibinfo  {journal} {Physical Review B}\ }\textbf {\bibinfo
  {volume} {41}},\ \bibinfo {pages} {7892} (\bibinfo {year}
  {1990})}\BibitemShut {NoStop}%
\bibitem [{\citenamefont {Pizzi}\ \emph {et~al.}(2016)\citenamefont {Pizzi},
  \citenamefont {Cepellotti}, \citenamefont {Sabatini}, \citenamefont
  {Marzari},\ and\ \citenamefont {Kozinsky}}]{pizzi_aiida_2016}%
  \BibitemOpen
  \bibfield  {author} {\bibinfo {author} {\bibfnamefont {G.}~\bibnamefont
  {Pizzi}}, \bibinfo {author} {\bibfnamefont {A.}~\bibnamefont {Cepellotti}},
  \bibinfo {author} {\bibfnamefont {R.}~\bibnamefont {Sabatini}}, \bibinfo
  {author} {\bibfnamefont {N.}~\bibnamefont {Marzari}}, \ and\ \bibinfo
  {author} {\bibfnamefont {B.}~\bibnamefont {Kozinsky}},\ }\href {\doibase
  10.1016/j.commatsci.2015.09.013} {\bibfield  {journal} {\bibinfo  {journal}
  {Computational Materials Science}\ }\textbf {\bibinfo {volume} {111}},\
  \bibinfo {pages} {218} (\bibinfo {year} {2016})}\BibitemShut {NoStop}%
\bibitem [{\citenamefont {Huber}\ \emph {et~al.}(2020)\citenamefont {Huber},
  \citenamefont {Zoupanos}, \citenamefont {Uhrin}, \citenamefont {Talirz},
  \citenamefont {Kahle}, \citenamefont {H{\"a}uselmann}, \citenamefont
  {Gresch}, \citenamefont {M{\"u}ller}, \citenamefont {Yakutovich},
  \citenamefont {Andersen}, \citenamefont {Ramirez}, \citenamefont {Adorf},
  \citenamefont {Gargiulo}, \citenamefont {Kumbhar}, \citenamefont {Passaro},
  \citenamefont {Johnston}, \citenamefont {Merkys}, \citenamefont {Cepellotti},
  \citenamefont {Mounet}, \citenamefont {Marzari}, \citenamefont {Kozinsky},\
  and\ \citenamefont {Pizzi}}]{huber_aiida_2020}%
  \BibitemOpen
  \bibfield  {author} {\bibinfo {author} {\bibfnamefont {S.~P.}\ \bibnamefont
  {Huber}}, \bibinfo {author} {\bibfnamefont {S.}~\bibnamefont {Zoupanos}},
  \bibinfo {author} {\bibfnamefont {M.}~\bibnamefont {Uhrin}}, \bibinfo
  {author} {\bibfnamefont {L.}~\bibnamefont {Talirz}}, \bibinfo {author}
  {\bibfnamefont {L.}~\bibnamefont {Kahle}}, \bibinfo {author} {\bibfnamefont
  {R.}~\bibnamefont {H{\"a}uselmann}}, \bibinfo {author} {\bibfnamefont
  {D.}~\bibnamefont {Gresch}}, \bibinfo {author} {\bibfnamefont
  {T.}~\bibnamefont {M{\"u}ller}}, \bibinfo {author} {\bibfnamefont {A.~V.}\
  \bibnamefont {Yakutovich}}, \bibinfo {author} {\bibfnamefont {C.~W.}\
  \bibnamefont {Andersen}}, \bibinfo {author} {\bibfnamefont {F.~F.}\
  \bibnamefont {Ramirez}}, \bibinfo {author} {\bibfnamefont {C.~S.}\
  \bibnamefont {Adorf}}, \bibinfo {author} {\bibfnamefont {F.}~\bibnamefont
  {Gargiulo}}, \bibinfo {author} {\bibfnamefont {S.}~\bibnamefont {Kumbhar}},
  \bibinfo {author} {\bibfnamefont {E.}~\bibnamefont {Passaro}}, \bibinfo
  {author} {\bibfnamefont {C.}~\bibnamefont {Johnston}}, \bibinfo {author}
  {\bibfnamefont {A.}~\bibnamefont {Merkys}}, \bibinfo {author} {\bibfnamefont
  {A.}~\bibnamefont {Cepellotti}}, \bibinfo {author} {\bibfnamefont
  {N.}~\bibnamefont {Mounet}}, \bibinfo {author} {\bibfnamefont
  {N.}~\bibnamefont {Marzari}}, \bibinfo {author} {\bibfnamefont
  {B.}~\bibnamefont {Kozinsky}}, \ and\ \bibinfo {author} {\bibfnamefont
  {G.}~\bibnamefont {Pizzi}},\ }\href {\doibase 10.1038/s41597-020-00638-4}
  {\bibfield  {journal} {\bibinfo  {journal} {Scientific Data}\ }\textbf
  {\bibinfo {volume} {7}},\ \bibinfo {pages} {300} (\bibinfo {year}
  {2020})}\BibitemShut {NoStop}%
\end{thebibliography}%

\end{document}